\documentclass[aps,prl,amssymb,showpacs,twocolumn]{revtex4}
\usepackage{graphicx}

\begin{document}

\title{CSL-1: Lensing by a Cosmic String or a Dark Matter Filament?}

\author{Malcolm Fairbairn}
\email{malc@physto.se}
\affiliation{Cosmology, Particle astrophysics and String theory, Department of Physics, Stockholm University, AlbaNova University Centre, SE-106 91, Stockholm, Sweden}


\newcommand{\lP}{\ell_{\mathrm P}}

\newcommand{\md}{{\mathrm{d}}}
\newcommand{\Kern}{\mathop{\mathrm{ker}}}
\newcommand{\tr}{\mathop{\mathrm{tr}}}
\newcommand{\sgn}{\mathop{\mathrm{sgn}}}

\newcommand*{\R}{{\mathbb R}}
\newcommand*{\N}{{\mathbb N}}
\newcommand*{\Z}{{\mathbb Z}}
\newcommand*{\Q}{{\mathbb Q}}
\newcommand*{\C}{{\mathbb C}}

\begin{abstract}
The lens candidate CSL-1 has been interpreted as evidence for a cosmic string. Here we test the hypothesis that the lensing comes from a tidally disrupted dark matter halo.  We calculate the mass-density relationship that one would expect from structure formation theory and come to the conclusion that in order to explain the lensing using dark matter, the halo would have to have a mass greater than the Milky Way.  There is apparently no such object seen in the data.  If the follow up observations confirm that the two objects are indeed images of the same galaxy, then it seems difficult to explain the lens using dark matter.
\end{abstract}
\maketitle

The Capodimonte-Sternberg Lens candidate 1 (CSL-1) consists of two objects with identical spectra and redshifts ($z=0.46$) separated by an angular distance of about 1.8''=$8.6\times 10^{-6}$ radians \cite{sazhin}.  The spectra are consistent with giant elliptical galaxies, and if they were to lie at precisely the same redshift, then their maximum separation would be around 11 kpc (assuming $\Omega_m$=0.3, $\Omega_\Lambda$=0.7, h=0.65 for the cosmological parameters).

It is possible that these two images are two separate galaxies located along the same line of sight with similar observed redshifts.  However, the similarity between their spectra has lead Sazhin{\it et al} to suggest that the two objects may both be images of the same object viewed through a gravitational lens \cite{sazhin}. However, there is no visible lensing candidate in the same field and since both images have very similar apparently spherical morphologies, a spherical lensing candidate is disfavoured.  For this reason it has been suggested that the gravitational lens may in fact be a cosmic string.  

Cosmic strings are classical solutions of field theory lagrangians which occur when there is a symmetry breaking from a symmetric state to a vacuum state with a U(1) degeneracy. They are predicted in grand unified extensions of the standard model \cite{gutstring} and are also generic in string theories \cite{witten,polchinski}. 

When the field theory sector is coupled to gravity, the cosmic string gives rise to a conical singularity so that an observer circumnavigating the string at a radius $r$ will get back to where she or he started after traveling less than $2\pi r$, in other words there is a deficit angle \cite{grstring}.  Light passing a cosmic string will be deflected by this angle and in this way, one can obtain multiple images of the same object.

It has been pointed out that if the lensing is due to the deficit angle of a cosmic string, then the tension associated with the string would be close to the tension one would expect from a symmetry broken at the GUT scale \cite{sazhin}.

The purpose of this letter is to investigate the hypothesis that the dual images may be due to the gravitational lensing effect of a tube of dark matter rather than a cosmic string.

Theoretical predictions and N-body simulations both suggest that there should exist a spectrum of dark halos of different masses, not just galaxy masses \cite{stefan,diemand}.  In fact, the number of sub-halos that one might expect in the region of the milky way is much larger than the number of Magellanic cloud-like objects observed (see \cite{hayashi} and references therein).  The way it is thought the cold dark matter paradigm circumvents this problem is the following: Halos with masses much less than a galaxy will have a less deep potential well and shocks created inside those halos due to supernovae or re-ionisation may blow baryons out of the halo \cite{bullblow}.

Some WIMP halos will be tidally stretched into filaments.  The distortion of the Sagittarius dwarf galaxy into a tidal stream \cite{sagit} and the characteristic filaments thrown off during the Antennae galaxy collisions \cite{antennae} are the two most obvious example of this phenomenon.  It is therefore interesting to see if one could use such filaments to explain the lensing event.  First we will work out the maximum radius of the halo from geometry.  This will lead to a minimum density in order to obtain the correct lensing.  Then we will use structure formation theory to obtain the mass-density relationship between halo density and masses.  Finally we will bring the results together and dicuss their implications.
 
\section{Maximum radius of the filament\label{maxrad}}
\begin{figure}[h]
\begin{center}
\includegraphics[height=5cm,width=8cm]{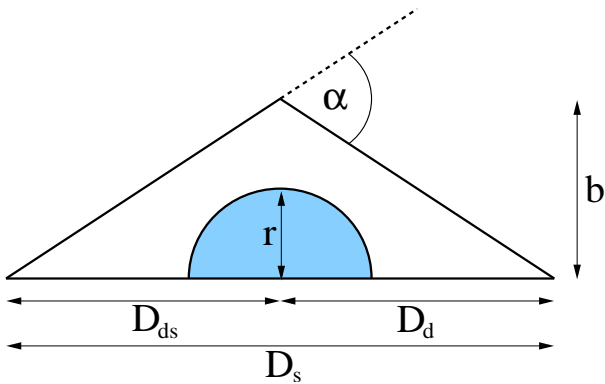} 
\caption{\it Lensing Geometry. The shaded region corresponds to the cylinder of dark matter seen long its length.}
\label{lens}
\end{center}
\end{figure}

The lensing geometry that we will consider is portrayed in figure \ref{lens}.  
In an expanding universe in order to obtain the impact parameter $b(z)$, we should evaluate the expression for the angular distance, i.e.
\begin{equation}
\frac{b}{{\rm Mpc}}=\theta d_A(z)=\frac{3\times 10^3\theta}{h(1+z)}\int_0^z\frac{dz'}{\sqrt{\Omega_\Lambda+\Omega_m(1+z')^3}}
\end{equation}
where $h=H_0/100 {\rm kms^{-1}Mpc^{-1}}$ (actually, the small angle approximation $b=\alpha D_d$ is good for redshifts $z\le 0.46$).
\begin{figure}[b]
\begin{center}
\includegraphics[height=9cm,width=7cm,angle=270]{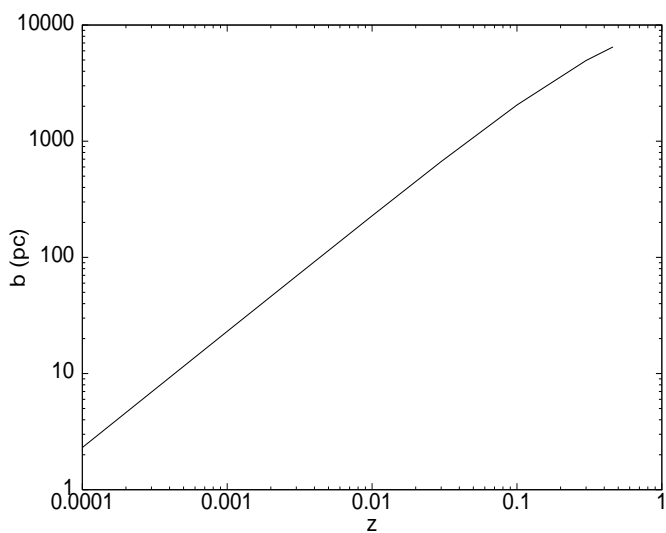} 
\caption{\it The impact parameter $b$ in parsecs as a function of redshift with $\Omega_m=0.3$, $\Omega_\Lambda=0.7$ and $h=0.65$.}
\label{angdist}
\end{center}
\end{figure}
Figure \ref{angdist} shows us that the maximum distance between the two photon paths leads to the constraint
\begin{equation}
r\le 6000 {\rm pc}
\end{equation}
so it will be necessary to place enough density inside a tube of that radius to obtain enough lensing.  Next we will calculate that density by looking at the equations for lensing by a filament.
\section{Minimum density of the filament}
 If we assume that the dark matter tube has a Gaussian density distribution of width $r$ and central density $\rho$ then the lensing angle for a photon passing at a distance $b$ from the centre of the tube is given by \cite{bozza} (Newton's constant $G=4.785\times 10^{-14} pc M_\odot^{-1}$)
\begin{equation}
\alpha=G\mu {\rm Erf}\left(\frac{b}{r\sqrt{2}}\right) 
\label{boz}
\end{equation}
where $\mu\simeq\pi r^2\rho$ is the mass per unit length of the tube. This equation is only valid when the length of the tube $l\gg b$.  

Equating $G\mu$ with the lensing angle $\theta$ and assuming the relationship $b\ge r$ then we have an expression for the minimum density of the object
\begin{equation}
\rho>\rho_{min}=\frac{\mu}{\pi r_{max}^2}= 0.8 {\rm M_\odot pc^{-3}}
\end{equation}
and now we need to find out if this density is reasonable in terms of the densities that we would expect for WIMP halos as calculated from the power spectrum.
\section{Maximum density/Minimum radius of the filament}
To find out the maximum possible density of a candidate filament, we will have to make a key assumption.  We are suggesting that a dark matter halo may be disrupted into a stream of matter during a tidal encounter with another halo.  Since we have no indication of exactly what kind of interaction that may be, we will have to assume that the typical density of the tube of dark matter is approximately the same as the halo which was disrupted to create the tube.  

Having made this assumption, then we should calculate the typical densities that we expect dark matter halos to have.  In order to do this, we have to take into account that smaller WIMP halos typically form at earlier times and have larger densities corresponding to the higher average density of matter during those epochs.  We will calculate the mass-density relationship for halos using the toy model presented by Bullock et al \cite{bullock}.  In this model, the typical virial mass $M_{vir}$ of halos which collapse at a redshift $z_c$ is the mass which solves the equation
\begin{equation}
\sigma(F M_{vir}(z_c))=\delta_{coll}(z_c)=\frac{1.686}{(1+z_c)}
\label{F}
\end{equation}
where $F$ is a parameter derived from N-body simulations and $\delta_{coll}$ is the critical over-density for the collapse of a spherical shell in a matter dominated universe.  Likewise the characteristic density $\rho_{char}$ of a dark matter halo which collapses at a redshift $z_c$ is given by
\begin{equation}
\rho_{char}(z_c)=K^3 \Delta_{vir}(z_c)\frac{3H_0^2}{8\pi G}\Omega_{m0} (1+z_c)^3\end{equation}
where $\Delta_{vir}$ is given for a flat universe by \cite{ullio}
\begin{equation}
\Delta_{vir}(x)\simeq \frac{18\pi^2+82x-39x^2}{1-x}
\end{equation}
and $x=\Omega_m(z)-1$. Since we will be considering halos which collapse at redshifts $z_c\gg 1$ where radiation and acceleration are irrelevant, we will assume that $\Delta_{vir}=178$, then following the fits of \cite{bullock} for simulations of halos  in a $\Lambda CDM$ universe we assume that $K=4$ so that 
\begin{equation}
\rho_{char}(z_c)=3.16\times 10^{-3}\Omega_{m0} h^2 (1+z_c)^3 {\rm M_{\odot}\ pc^{-3}}
\end{equation}
so that now we must work out the rms magnitude of perturbations for different mass scales $\sigma(M)$.

\begin{figure}[h]
\begin{center}
\includegraphics[height=7cm,width=9cm]{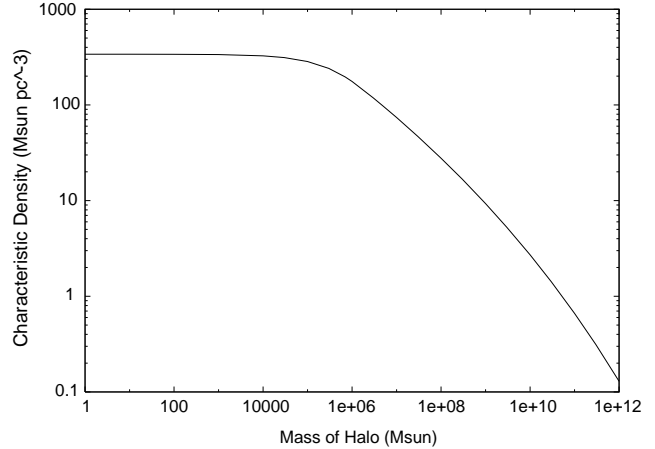} 
\caption{\it Density-mass relation for halos.  The density of less massive halos increases with a maximum value around 100 M$_\odot$ pc$^{-3}$ }
\label{densfig}
\end{center}
\end{figure}

In order to obtain $\sigma(M)$, we need to assume an adiabatic CDM evolution for the perturbations, so that the transfer function between the primordial fluctuations and those at late times goes like
\begin{eqnarray}
T(k)=\frac{\ln(1+2.34q)}{2.34q}\left[1+3.89q+(16.1q)^2 \right.\nonumber\\
\left. +(5.46q)^3+(6.71q)^4\right]^{-1/4}
\end{eqnarray}
where $q=k{\rm Mpc}/\Omega_m h$ when the effect of baryons is not taken into account \cite{bardeen86}. Including baryons we use a version of $q$ which takes into account baryonic suppression on small scales \cite{sugiyama95}
\begin{equation}
q=\frac{k {\rm Mpc}}{\Omega_m h \exp\left[-\Omega_b(1+\sqrt{2h}/\Omega_m)\right]}.
\end{equation}
We define $\sigma(M)$ to be the rms density fluctuation on the scale encompassing a mass $M$, given by
\begin{equation}
\sigma^2(M)=\int d^3k \tilde{W}^2(k R)P(k) 
\end{equation}
where $\tilde{W}$ is the top hat window function and $R^3=3M/4\pi\Omega_m \rho_{crit}$ where $\rho_{crit}=2.78\times 10^{11} M_\odot {\rm Mpc^{-3}}$. The spectrum is normalised by setting
\begin{equation}
\sigma_8=\sigma\left(\frac{0.173}{h{\rm Mpc}}\right)=0.897
\end{equation}
then if we set $h=0.65$ we find that $\sigma_8=\sigma(1.53\times10^{13}{\rm M_\odot})$.  Again referring to the numerical fits of \cite{bullock} we take the fraction in equation (\ref{F}) to be given by $F=0.01$ and then we obtain the results plotted in figure \ref{densfig}.
The results in figure 3 tell us that more massive halos have a lower characteristic density and that lower mass halos all have a similar characteristic density, around $\rho_{max}=$300 M$_\odot$pc$^{-3}$.  This gives us our maximum density for the lensing object
\begin{equation}
\rho < \rho_{max}=300 {\rm M_\odot pc^{-3}}
\end{equation}
Equation (\ref{boz}) then tells us exactly what the radius of the tube needs to be for a given density, so that we now have a minimum radius for the tube.
\begin{equation}
r_{min} > \sqrt{\frac{2\pi \alpha}{G\rho_{max}}}\simeq 200 {\rm pc}
\label{rmin}
\end{equation}
So that structure formation theory has given us an upper limit on the typical density (and it {\it is} greater than the lower limit) that we might expect for a WIMP halo, and therefore a minimum radius and mass that we would require if we were to try and use such a halo as our lens candidate.
\section{CSL-1 as dark matter halo?}
In order to find out if the density predicted by structure formation theory is large enough to explain the lensing event we have made some assumptions.  One should of course perform computer simulations to find out how the density changes before and after the halos are stretched into filaments via tidal distortion. Here, as a first approximation, we will assume that the density of the filament of dark matter is equal to the density of the halo with the same total mass as the filament.  This will probably overestimate the density of the filament.

Next we will assume that the length of the tube is twice its diameter, i.e. $l=ar$, $a=4$ this seems the minimum assumption that we would require since the result (\ref{boz}) is valid only if the impact parameter of the photons $b\ll l$.

\begin{figure}[t]
\begin{center}
\includegraphics[height=8.5cm,width=6.5cm,angle=270]{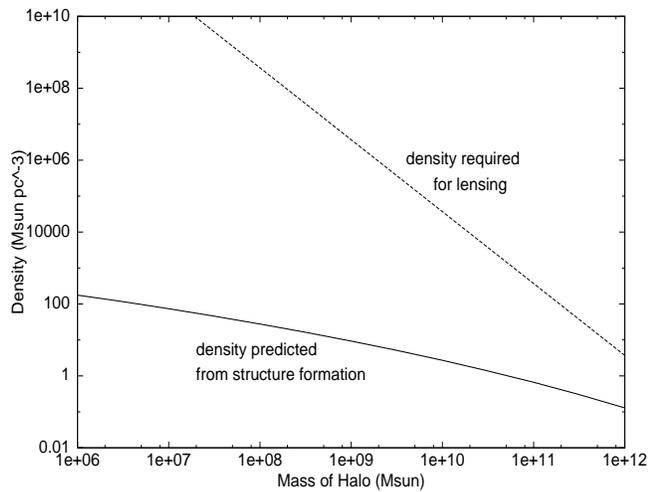} 
\caption{\it Density predicted by structure formation theory vs. density required to explain the lensing event.}
\label{compare}
\end{center}
\end{figure}

We can calculate the density that would be required in order to explain the lensing using $\alpha=G\mu=G\rho \pi r^2$ and $m=\rho \pi r^2 a r$.  The required density is then given by
\begin{equation}
\rho_{required}=\frac{a^2\alpha^2}{\pi m^2 G^3}
\end{equation}
and we plot $\rho_{required}$ as a function of mass in figure \ref{compare} vs. the density-mass relation for halos we get from structure formation (we know the lensing angle $\alpha$ from the observation).
Figure \ref{compare} shows us that in order to try and explain the lensing via a WIMP halo, we should consider higher masses.  This is at first perhaps slightly surprising, since we have shown that the density of WIMP halo {\it decreases} at higher densities, but a moment's thought tells us that $\rho r^2\sim\rho^{1/3}m^{2/3}$, so the dependence on density is actually weak.  It is therefore clear from figure \ref{compare} that in order to explain the lensing using a WIMP halo, we would have to consider one with mass $M_{halo}\ge 10^{12}M_\odot\sim M_{MilkyWay}$.  Since it is difficult to imagine how such a halo could exist without having significant numbers of baryons in it, these results make it difficult to reconcile the non-observation of a lensing candidate in the data \cite{sazhin} with the idea of dark matter being responsible for the lensing.

There are two caveats to this logic which should be mentioned here.  The first is that the most up to date simulations of dark matter halos show that they have a very cuspy centre, and unlike the Gaussian we used in equation \ref{boz} , their inner profiles go like $\rho\propto r^{-\gamma}$ with $\gamma= 1.2$ \cite{1.18}.  In principle this effect could be used if the impact parameter $b$ was much less than the cross sectional radius of the tube $r$, although it seems unlikely that the central density of the dark matter tube would remain as cuspy after tidal disruption.

The second possibility is the use of dark matter caustics which can come in various shapes including tubes.  However, more detailed analysis shows that it is unlikely caustics can create an observable signal, even at high redshifts \cite{charmousis} (although see \cite{onemli}).

In conclusion, we have tested the hypothesis that the CSL-1 lensing candidate could be explained via lensing due to a tidally disrupted halo of dark matter.  We have shown that such a halo would probably have to have a mass bigger than the Milky Way, and since we see no such object in the CSL-1 field, we can neglect this possibility.  The simplest explanation for the observation is of course that the two images are two separate galaxies at the same redshift\footnote{The author understands that more detailed observations are being made with the Hubble Space telescope.}.  If this turns out to not be the case, we will certainly have an intriguing mystery on our hands.

{\bf Acknowledgments} I am grateful for conversations with Phillipe Brax, Joakim Edsj\"o and James Gray and for funding from the Swedish Research Council (Vetenskapsr\aa det).

\end{document}